\documentclass[page-classic]{epl2} 

\usepackage{amsmath}
\usepackage{color}
\usepackage{ulem}

\newcommand{\marge}[1]{\marginpar{}}  
\newcommand{\Sl}[1]{{}}           
\newcommand{\beq}[1]{\Sl{#1}\begin{equation}\if#1\empty\else\label{#1}\fi}
\newcommand{\eeq}{\end{equation}}
\newcommand{\beqa}[1]{\Sl{#1}\begin{eqnarray}\if#1\empty\else\label{#1}\fi}
\newcommand{\eeqa}{\end{eqnarray}}

\newcommand{\nm}{\nonumber\\}
\newcommand{\eq}[1]{(\ref{#1})}

\newcommand{\la}{\langle}
\newcommand{\ra}{\rangle}

\newcommand{\alphp}{\alpha^\prime}
\newcommand{\betap}{\beta^\prime}
\newcommand{\alphapp}{\alpha\prime\prime}

\shorttitle{Nonextensive formalism and continuous Hamiltonian systems}
\institute{
 \inst{1} Center for Nonlinear Phenomena and Complex Systems CP 231 \\
  Universit\'{e} Libre de Bruxelles, 1050 Brussels, Belgium 
}

\begin{document}

\title{ Nonextensive formalism and \\
continuous Hamiltonian systems}
\author{{Jean Pierre Boon$^1$\thanks{E-mail: \email{jpboon@ulb.ac.be}} and
James F. Lutsko$^1$\thanks{E-mail: \email{jlutsko@ulb.ac.be}}}}
\shortauthor{J.P. Boon and J.F. Lutsko}



\abstract{
The homogeneous entropy for continuous systems in nonextensive statistics 
reads $S^{H}_{q}=k_B\,{(1 - (K \int d\Gamma \rho^{1/q}(\Gamma))^{q} )}/({1-q})$, 
where $\Gamma$ is the phase space variable.
Optimization of  $S^{H}_{q}$ combined with normalization and energy 
constraints gives an implicit expression of the distribution function 
$\rho (\Gamma)$ which can be computed explicitly when the Hamiltonian
reduces to its kinetic part. We examine the meaning of the $q$-{\it ideal gas} and 
we compute its properties such as the energy fluctuations and the specific heat. 
Similar results are also presented using the formulation based on the  Tsallis  entropy. 
From the analysis, we find that the validity of the nonextensive formalism 
for the $q$-{\it ideal gas} is restricted to the range $q<1$, which raises the question
of the formal validity range for continuous Hamiltonian systems.}
 
\maketitle

\section{Introduction}

Many phenomena in natural systems and in laboratory experiments are observed and 
measured under non-equilibrium conditions, and therefore do not obey the standard statistical 
mechanics  description. In particular the distributions which characterize such systems  are not
Boltzmann-like and do not follow from linear response. Instead these distributions exhibit "fat tails" 
and power law decays and often they can be fitted 
by $q$-exponential functions which generalize the usual Boltzmann exponential distribution
\cite{swinney-tsallis}. A recurring question is what deviation from standard statistical mechanics 
gives rise to this behavior, which amounts
to the question of the emergence of "statistics from dynamics" as emphasized by E.G.D. Cohen
\cite{cohen}. There are several possible analytical developments from which  
$q$-exponential distributions can be obtained: superstatsitics \cite{beck-cohen} by statistical  average 
over the $\chi$-square distribution of an intensive variable, nonlinear response theory \cite{lutsko-boon} 
by the solution of the generalized Fokker-Planck equation, and nonextensive statistics \cite{Tsallis88} 
by optimization of the generalized entropy.

It was precisely  the original idea of nonextensive statistics introduced by Tsallis about
20 years ago\cite{Tsallis88} to develop a statistical mechanical theory for systems out of 
equilibrium where the Boltzmann distribution no longer holds, and to generalize the Boltzmann 
entropy by a more general function $S_q$ while maintaining the formalism of thermodynamics.
From a practical viewpoint, the nonextensive statistics formulation appeared
to be of interest because maximization of the generalized entropy under the 
usual constraints (normalized probabilities, fixed internal energy) yields the
$q$-exponential distribution {which has been successfully used to describe distributions 
observed  in a large class of phenomena \cite{Tsallis09}. Indeed for a certain range of values 
of the index $q$, these $q$-exponential distributions exhibit a power law decay (when $q > 1$),
a feature observed in a large class of experimental phenomena 
which cannot be straightforwardly interpreted in the context of classical theories.
{At the same time, a large literature concerning the internal self-consistency of the nonextensive formalism has developed addressing such questions as the stability of the entropy functional \cite {lutsko_boon_grosfils}, the method of calculating averages \cite{abe_99a} and the positivity of the specific heat \cite{abe_99b, feng}.}

Many explicit applications of the nonextensive formalism involve the assumption of independent particles: e.g., non-interacting particles that can occupy a set of discrete energy levels or, in the classical case, the ideal gas. These applications may appear paradoxical as the assumption of non-extensivity implies an interaction between components of the system. An alternative point of view, adopted here, is that there is no paradox because the underlying physical system does involve interactions (even long-ranged interactions). The independent particles are not the constituents of the physical system, but rather are understood to be quasi-particles in which the effect of the interactions are, to a first approximation, included in their properties (effective mass, statistics, etc.). With this point of view, the adoption of the nonextensive formalism is another part of the effective one-body description required to account for aspects of the interactions that cannot be otherwise modeled.} Accordingly we revisit the formalism for the $q$-{\it ideal gas}.

 Even this simple case of the $q$-ideal gas has been subject to questions of internal consistency. In particular, that the nonextensive formalism has limited range of validity \cite{abe_99a}, gives negative specific heats \cite{abe_99b} and even, recently, negative values for the second cumulant of the energy (a positively defined quantity!)\cite{feng}.
Here we re-examine these issues using both the Tsallis entropy and, as suggested recently \cite{lutsko_boon_grosfils}, the homogeneous entropy.

\section{The homogeneous entropy}

We start with the homogeneous entropy (or normalized Tsallis entropy) 
$S^{H}_{q}$  which was proven to be stable against small perturbations
in the probability distribution function $\rho (\Gamma)$ while the Tsallis entropy 
$S_q$ is not  \cite{lutsko_boon_grosfils}. For continuous systems the 
$H$-entropy is given by
\begin{equation}
  S^{H}_{q}= k_B\,\frac{1 - (K \int d\Gamma \rho^{1/q}(\Gamma))^{q}}{1-q} \,,
\label{S_H}
\eeq
where  $q$ is the index characterizing the entropy functional, $\Gamma$ denotes 
the phase space variable, and $K$ must be a quantity 
with the dimensions of $\left[ \Gamma \right] ^{\frac{1-q}{q}}$, i.e. 
$K= \hbar ^{ND\left( {\frac{1-q}{q}}\right) }$ with $N$, the number of degrees of 
freedom of the system with dimension $D$ and Hamiltonian $H$.
{In the limit  $q \rightarrow 1$,  the classical Boltzmann-Gibbs formulation is retrieved.}%
\footnote{For simplicity the Boltzmann factor $k_B$  will be omitted and 
reincluded explicitly when necessary.}

Optimization of the $H$-entropy (\ref{S_H}) with the normalization and energy constraints:%
\begin{equation}
1 \,=\,\int \rho \left( \Gamma \right) d\Gamma \;\;\;\;\;;\;\;\;\;\;\;
U \,=\,{\int \rho \left( \Gamma \right) H d\Gamma } \,, \label{U_cons}
\end{equation}%
by the method of Lagrange multipliers  leads to 
\begin{eqnarray}
0 &=&\frac{\delta }{\delta \rho \left( \Gamma \right) }\left(S^{H}_{q}  - \alpha
\left( \int \rho \left( \Gamma \right) d\Gamma -1\right) -\beta \left( \int
\rho \left( \Gamma \right) Hd\Gamma -U\right) \right) \nm
&=& \frac{-q\left( K\int \rho ^{1/q}\left( \Gamma \right) d\Gamma \right)
^{q-1}K\frac{1}{q}\rho ^{\frac{1}{q}-1}\left( \Gamma \right) }{1-q}-\alpha - \beta H  \,,
\label{opt_SH}
\end{eqnarray}%
which is solved to give%
\begin{eqnarray}
\rho \left( \Gamma \right) &=& \left( \frac{\left( q-1\right) \alpha }{\left(
K\int \rho ^{1/q}\left( \Gamma \right) d\Gamma \right) ^{q-1}K}+\frac{\left(
q-1\right) \beta }{\left( K\int \rho ^{1/q}\left( \Gamma \right) d\Gamma
\right) ^{q-1}K}H\right) ^{\frac{q}{1-q}} \nonumber\\
&=& {\cal Z}^q_q \,\left(\alpha^\prime + \beta^\prime H \right)_+^{\frac{q}{1-q}} \,,
\label{rhoH}
\end{eqnarray}%
where ${\cal Z}_q = K \int \rho ^{1/q}\left( \Gamma \right) d\Gamma$,
and $\alpha^\prime = (q-1) \alpha /K$ and $ \beta^\prime = (q-1) \beta /K$.

$\rho \left( \Gamma \right) $ is the physical probability distribution,
which must be real, positive and normalizable; so we must have
\begin{equation}
1 =  {\cal Z}^q_q  \int \,\left(\alpha^\prime + \beta^\prime H \right)_+^{\frac{q}{1-q}}%
d\Gamma \,.\label{normH}%
\end{equation}
We will also consider the first moments of the Hamiltonian
\begin{equation}
 \la H^m \ra =  \int \rho \left( \Gamma \right) H^m d\Gamma \,=\,
 {\cal Z}^q_q  \int \,\left(\alpha^\prime + \beta^\prime H \right)_+^{\frac{q}{1-q}}%
H^m d\Gamma \,;\label{Hm}%
\end{equation}
in particular we are interested in 
\begin{equation}%
U =  \la H \ra = \frac{\int \,\left(\alpha^\prime + \beta^\prime H \right)^{\frac{q}{1-q}}%
H d\Gamma}{\int \,\left(\alpha^\prime + \beta^\prime H \right)^{\frac{q}{1-q}}%
d\Gamma} \;\;\;\;\;;\;\;\;\;\;\;
\la H^2 \ra = \frac{\int \,\left(\alpha^\prime + \beta^\prime H \right)^{\frac{q}{1-q}}%
H^2 d\Gamma}{\int \,\left(\alpha^\prime + \beta^\prime H \right)^{\frac{q}{1-q}}%
d\Gamma}\,,
\label{U-H2}
\end{equation}%
where we used the normalization condition in (\ref{U_cons}).
So the integrals to be considered have the form
\begin{equation}
 I_m  =  \int \,\left(\alpha^\prime + \beta^\prime H \right)_+^{\frac{q}{1-q}}%
H^m d\Gamma \,,\label{Im}%
\end{equation}
and the result of the integration will depend on the sign of the Lagrange multipliers; 
therefore we must consider the following possible cases:  (i)  $\alphp>0$ and $\betap<0$,
{(ii) $\alphp>0$ and $\betap>0$, and (iii) $\alphp<0$ and $\betap>0$}
(If both $\alphp, \betap<0$, there is no solution to the normalization condition (\ref{normH})).

\section{The $q$-ideal gas}

Besides the physical meaning  of the $q$-ideal gas which was 
explained in the introductory section, it is legitimate to discuss its validity  
in the context of the nonextensive  formalism because, if the formalism 
is to be used for continuous Hamiltonian systems, it should first pass the 
test of the $q$-ideal gas (as in classical statistical mechanics).

For the $q$-ideal gas, the Hamiltonian reduces to its kinetic part and the configuration integral in (\ref{Im})
is straightforward and yields a factor given by the space volume $V^N$. With a change of variable
$X=\frac{\betap}{\alphp}\,\frac{p^2}{2m}$, (\ref{Im}) for $\alphp>0$ and $\betap<0$
{(case (i))} becomes
\begin{eqnarray}
I^{IG}_m&=&V^{N} S_{DN} \left(2m\right)^{ND/2} \frac{1}{2} {\alphp}^{\frac{q}{1-q}}
\left(\frac{\alphp}{|\betap|}\right)^{\frac{ND}{2}+m}\int
_{0}^{\infty}\left(1-X\right)  ^{\frac{q}{1-q}}X^{\frac{ND}{2}%
+m -1}dX \nonumber\\
&=&V^{N} S_{DN} \left(2m\right)^{ND/2} \frac{1}{2} {\alphp}^{\frac{q}{1-q}}
\left(\frac{\alphp}{|\betap|}\right)^{\frac{ND}{2}+m}B\left(\frac{1}{1-q} ,%
\frac{ND}{2} + m\right)\,,
\label{I^IG}
\end{eqnarray}
where $B(k,l)$ is the Beta function
provided $q<1$ (i.e. $\alpha <0$, and  $\beta>0$), and excludes the possibility
$q>1$ (with $\alpha>0$ and  $\beta<0$). 
Then we have
\begin{equation}
\frac{I^{IG}_m}{I^{IG}_0} = \left( \frac{\alphp}{|\betap|}\right)^{m}%
\frac{B\left(\frac{1}{1-q},  \frac{ND}{2}+m\right)}%
{B\left(\frac{1}{1-q},  \frac{ND}{2}\right)  }\,,
\end{equation}
which, with (\ref{U-H2}), gives
\begin{equation}
\frac{\alphp}{|\betap|}  = U\left(  1 + \frac{2}{(1-q)ND}\right) \,,%
\end{equation}
and
\begin{eqnarray}
\left\langle H^{2}\right\rangle -\left\langle H\right\rangle ^{2} &=& \left(
\frac{\alphp}{|\betap|}\right)  ^{2}\frac{B\left(  \frac{1}{1-q},\frac{ND}\,
{2}+2\right)  }{B\left(  \frac{1}{1-q},\frac{ND}{2}\right)  }-U^{2}\nonumber\\
&=& \frac{4 \; U^{2}}{ND \left( 2 + \left(  1-q\right) \left( 2+ ND \right)\right)  }\,.
\label{E_fluc}
\end{eqnarray}
Note that $\left\langle H^{2}\right\rangle -\left\langle H\right\rangle ^{2}$ is always positive  since the $q$ index must be $q<1$. It  also follows from these results that the explicit expression of the distribution function for the $q$-ideal gas is given by
\begin{equation}
\rho^{IG} \left(\Gamma \right) = \left( \frac{{\cal Z}_q}{K} \right)^{\frac{q}{1-q}}%
\left(1-(1-q)\frac{\beta}{{\cal Z}_q^{q}}\,(H-U)\right)^{\frac{q}{1-q}} \,,
\label{rhoH2}
\end{equation}
or, with the notation $\exp_q = (1 + (1-q) x)_+^{\frac{1}{1-q}}$, 
\begin{equation}
\rho^{IG} \left(\Gamma \right) = 
\frac{\left(\exp_q\frac{-\beta}{{\cal Z}_q^{q}}\,(H-U)\right)^{q}}%
{\int \exp_q\frac{-\beta}{{\cal Z}_q^{q}}\,(H-U) \,d\Gamma}\,.
\end{equation} 
Noting that ${\cal Z}_{q=1}\,=\,1$, it is clear that for $q=1$, one retrieves the classical exponential distribution.

Proceeding along the same lines for {case  (ii):} $\alphp, \betap>0$, i.e. 
$\alpha, \beta>0$ with $q>1$, or $\alpha, \beta<0$ with $q<1$, we obtain 
\begin{equation}
\frac{I^{IG}_m}{I^{IG}_0} = \left(  \frac{\alpha}{\beta}\right)
^{m}\frac{B\left(  \frac{ND}{2}+m, \frac{q}{q-1}-(\frac{ND}{2}+m)\right)
}{B\left(  \frac{ND}{2}, \frac{q}{q-1}-\frac{ND}{2}\right)  }\,,
\end{equation}
if and only if $1<q<1+\frac{1}{\frac{ND}{2}+m - 1}$. This gives
\begin{eqnarray}
 \frac{\alpha}{\beta}  
&=& U \left(  \frac{2}{ND\left(q-1\right)  }-1\right)\,,\nonumber\\
\la H^2 \ra - {\la H \ra}^2 &=&  \frac{4 \; U^{2}}%
{ND \left( 2 - \left( q - 1\right) \left( 2+ ND \right)\right)  }\,,
\label{H2_1<q}
\end{eqnarray}
which is valid (positive definite) when $1<q<1+\frac{2}{ND+2}$.
Notice that this range of the $q$ index is vanishingly small for $ND >> 1$ and therefore
physically negligible. In this case $\alpha, \beta>0$, but the case $\alpha, \beta<0$ 
(with $q<1$) is excluded.

 For $\alphp<0$ and $\betap>0$ {(case  (iii))}, we have 
\begin{equation}
I_m=V^{N} S_{DN} \left(2m\right)^{ND/2} \frac{1}{2} {|\alphp|}^{\frac{q}{1-q}}%
\left(\frac{|\alphp|}{\betap}\right)^{\frac{ND}{2}+m}\int%
_{0}^{\infty}\left( -1+X\right)  ^{\frac{q}{1-q}}X^{\frac{ND}{2}-1+m}dX \,,
\end{equation}
which, whether $q<1$ or $q>1$, has no solution. So the cases $q>1$ with $\alpha <0$, and  $\beta>0$,  and $q<1$ with $\alpha>0$, and  $\beta<0$ are excluded. 

In summary, we have shown that, except for the physically negligible range 
$1<q<1+\frac{2}{ND+2}$, the distribution function $\rho (\Gamma)$ for the $q$-ideal gas  is normalizable only for $q<1$ (with $\alpha <0$, and  $\beta>0$),
and that, contrary to some recent claim \cite{feng}, the positivity of the energy mean squared fluctuations $\left\langle \left(H  -\left\langle H\right\rangle \right)^{2} \right\rangle$ is always satisfied.

If, instead of the homogeneous entropy (\ref{S_H}), we start from the Tsallis entropy \cite{Tsallis88} for continuous systems 
$S_{q}=\frac{K \int d\Gamma \rho_T^{q}(\Gamma) - 1}{1 - q}$ , and use the same 
optimization procedure \eq{U_cons} (except that $U$ must then be computed with the escort average $U \,=\,\frac{{\int \rho_T^q \left( \Gamma \right) H d\Gamma }}{{\int \rho_T^q%
\left( \Gamma \right) d\Gamma }}$), we obtain the distribution function 
\begin{equation}
\rho_T \left( \Gamma \right) = \left(\alphapp -  \alphapp (1-q)\frac{\beta}{{\cal Z}_T} \,\left( H - U \right)\right) ^{\frac{1}{1-q}} \,,
\label{rhoT}
\end{equation}%
where $\alphapp = \frac{q}{1-q}\,\frac{K}{\alpha}$, and 
${\cal Z}_T = K \int \rho ^{q}\left( \Gamma \right)d\Gamma$.
Therefrom performing the computation for the $q$-ideal gas \cite{boon_lutsko} leads to conclusions 
that are the same as above and are in essential agreement with some results by Abe \cite{abe_99a, abe_99b};  in particular we find that  the normalized distribution function  exists only for $q<1$ (besides the physically vanishingly small (for $N>>1$) range $1<q<1+\frac{2}{ND+2}$)
with the additional observation that $\rho_T(\Gamma)$ has a singular point at $q=0$. 

\section{Thermodynamic quantities}

We now evaluate the homogeneous entropy starting from (\ref{S_H}) rewritten as
\begin{equation}
S^H_q = k_B \,\frac{1- {\cal Z}_q^{q}}{1-q}\,,%
\end{equation}
with
\begin{equation}
{\cal Z}_q^{q} = \frac{\left( K \int \left(\alpha^\prime + \beta^\prime H \right)_+^{\frac{1}{1-q}}%
d\Gamma \right)^q}{\int \left(\alpha^\prime + \beta^\prime H \right)_+^{\frac{q}{1-q}}%
d\Gamma}\,=\, K^q\, \frac{I^q}{I_0} \,.
\end{equation}
For the $q$-ideal gas with $q<1$ and $\beta > 0$, {using} (\ref{I^IG}), we find
\begin{equation}
{\cal Z}_q^{q} = K^q \left( V^{N} S_{DN} \left(2m\right)^{ND/2} \frac{1}{2}\right)^{q-1}%
\left(\frac{\alphp}{|\betap|}\right)^{\frac{ND}{2}(q-1)}\frac{\left(B\left(\frac{2-q}{1-q} ,%
\frac{ND}{2} \right)\right)^q}{B\left(\frac{2-q}{1-q}, \frac{ND}{2} \right)}\,,
\end{equation}
where
\begin{equation}
\frac{\left(B\left(\frac{2-q}{1-q} ,%
\frac{ND}{2} \right)\right)^q}{B\left(\frac{2-q}{1-q}, \frac{ND}{2} \right)}\,=\,
\frac {B^{q-1}\left(\frac{1}{1-q} ,\frac{ND}{2} \right)}{\left(1+(1-q)\frac{ND}{2}\right)^q} \,,
\end{equation}
and 
\begin{equation}
\frac{\alphp}{|\betap|}\,=\,U\left(1 + \frac {2}{(1-q) ND}\right)\,.
\end{equation}
Combining these results, we obtain
\begin{equation}
{\cal Z}_q^q\,=\, \frac{K^q \,R^H \left(  V;q\right)}{U^{\left(1-q\right)\frac{ND}{2}}}\,,%
\label{CalZ}
\end{equation}
with
\begin{equation}
R^H(V;q)\,=\,\left( V^{N} S_{DN} \left(2m\right)^{ND/2} \frac{1}{2}
\frac{\left(1+(1-q)\frac{ND}{2}\right)^{\frac{ND}{2}-\frac{q}{q-1}}}%
{\left((1-q)\frac{ND}{2}\right)^{\frac{ND}{2}}}\,B\left(\frac{1}{1-q} ,\frac{ND}{2} \right)%
\right)^{q-1}\,,%
\end{equation}
and
\begin{equation}
S^H_{q}=k_B\,\frac{1- K^q \,R^H \left(  V;q\right) \,U^{\frac{ND}{2}\left( q-1\right)}}{1-q}\,.%
\label{S^H_ID}
\end{equation}

It follows that the thermodynamic temperature of the $q$-ideal gas is given by
\begin{equation}
\frac{1}{T^H_q}\,=\, \frac{\partial S^H_{q}}{\partial U}= k_B\,K^q R^H\left(V;q\right)%
\frac{ND}{2}U^{\frac{ND}{2}\left(q-1\right)  -1} \label{THq} \,.
\end{equation}
In the limit $q\rightarrow 1$, $K = 1$ and $R^H\left(V;q\rightarrow 1\right) = 1$, 
so that for the classical ideal gas, where $ U= \frac{ND}{2} k_B T$, we retrieve the 
expression $\frac{\partial S}{\partial U}=  \frac{1}{T} $.
The specific heat is then readily obtained 
\begin{equation}
C^H_V\,=\,\left( \frac{\partial T^H_q}{\partial U}\right)^{-1}\,=\, k_B\,\frac{ND}{2}\,K^q\,R^H(V;q)\,\frac{U^{(q-1)\frac{ND}{2}}}{1+(1-q)\frac{ND}{2}}\;,
\label{Cv}
\end{equation}
which is always positive for $q<1$, and, for $q = 1$, gives the classical result
$C_V\,=\,  \frac{ND}{2}\,k_B$.  Note that using  \eq{E_fluc}, \eq{CalZ} and \eq{THq},
 \eq{Cv}  can also be written as
\begin{equation}
C^H_V = \frac{\left\langle H^{2}\right\rangle -\left\langle H\right\rangle ^{2}}{k_B\,(T^H_q)^{2}}
\,{\cal C}_q%
\end{equation}
with ${\cal C}_q\,=\,{\cal Z}_q^{-q}\,\left(1+\frac{1-q}{1+(1-q)\frac{ND}{2}}\right) $,
which generalizes the expression of the specific heat given in terms of the energy fluctuations 
$C_V =   {\la \left( \Delta E \right)^2 \ra}/\left({k_B\, T^2}\right)$. 

When we perform the same computation with the Tsallis formulation \cite{boon_lutsko}, 
we find the specific heat
\begin{equation}
C^q_V\,=\,k_B\frac{ND}{2}\,K\,R(V;q)\,\frac{U^{(1-q)\frac{ND}{2}}}{1-(1-q)\frac{ND}{2}}\,,
\end{equation}
which, in the limit  $q\rightarrow 1$, gives the classical expression for $C_V$, but where the denominator is {\it negative} for $q<1$, except when $q=1-\epsilon$ with 
{$\epsilon < \frac{2}{ND} << 1$}.  So, except in this narrow range, the Tsallis entropy 
formalism gives a negative specific heat  for the $q$-ideal gas.

\section{Concluding comments}

We have shown that optimization of the $H$-entropy {for continuous Hamiltonian 
systems} combined with normalization and energy constraints gives an expression for the 
distribution function which is computed explicitly for the  {$q$-}ideal gas and that, 
{in the thermodynamic limit,} the distribution function exists in the $q < 1$ index  range. 
We have also shown (i) that in this range the mean squared energy fluctuations are always positive,  {in contradition to recent claims that were a result of not taking into account the existence of intermediate integrals in the evaluation\cite{feng},} and (ii) that in the usual Tsallis formulation the specific heat of the $q$-ideal gas is negative for $q<1$.
We conclude that the use of the non-extensive formalism to "explain" observed $q$-exponential distributions on the basis of non-interacting quasi-particles is problematic when  $q > 1$, the range where the $q$-exponential function exhibits power law decay. 
Furthermore, in the range $q<1$ where the normalized distribution function exists, the Tsallis
formalism is also questionable as it gives a negative specific heat  for the {$q$-}ideal gas.
Its applicability to Hamiltonian systems with continuous canonical variables has also been questioned recently by Abe from a different viewpoint \cite{abe_09}.

As discussed in the introduction, there are two aspects to the nonextensive approach to the 
 study of nonequilibrium systems. (i) Nonextensive {\it statistics} has been applied succesfully to analyze 
 and to interpret observations in Hamiltonian systems which exhibit power law decay \cite{Tsallis09}; these  
 interpretations are based on phenomenological analyses in accordance with $q$-exponential distributions.
 (ii) The nonextensive {\it formalism} was constructed on the basis of a few axioms and accordingly should  
  develop  with self-consistency.  Our analysis suggests that the range of validity of the latter is limitted. It may be that there would be less restriction if the formalism were developed with interacting particles but in this case, the reason for assuming the nonextensive formalism becomes unclear. This however does 
  not preclude the pragmatic application of nonextensive statistics in phenomenological analyses of experimental results. 
 
\acknowledgments
This work was {partly} supported by the European Space Agency under contract
number ESA AO-2004-070.


\bigskip


\begin{thebibliography}{99}

\bibitem{swinney-tsallis}
  \Name{Swinney, H. L. \and Tsallis, C., eds}
  \REVIEW{Physica D, Anomalous distributions, Nonlinear dynamics, 
  and Nonextensivity}{193}{2004}{1-356}.
  
\bibitem{cohen}
  \Name{Cohen, E.G.D.}
  \REVIEW{Physica A}{305}{2002}{19}.  

\bibitem{beck-cohen}
  \Name{Beck, C. \and Cohen, E.G.D.}
  \REVIEW{Physica A}{321}{2003}{267}.
  
\bibitem{lutsko-boon}
  \Name{Lutsko, J.F. \and Boon, J.P.}
  \REVIEW{Phys. Rev. E}{77}{2008}{051103}.


\bibitem{Tsallis88}
  \Name{Tsallis, C.}
  \REVIEW{J. Stat. Phys.}{52}{1988}{479}.

\bibitem{Tsallis09}
  \Name{Tsallis, C.}
  \Book{Introduction to Nonextensive Statistical Mechanics}
  \Publ{Springer, New York}
  \Year{2009}.

\bibitem{lutsko_boon_grosfils}
  \Name{Lutsko, J.F., Boon,J.P. \and Grosfils, P.}
  \REVIEW{Europhys. Lett.}{86}{2009}{40005}.    
  
\bibitem{feng}
  \Name{Feng, Z.-H. \and Liu, L.-Y.}
  \REVIEW{Physica A}{389}{2010}{237}.
  
\bibitem{boon_lutsko}
  \Name{Boon,J.P. \and Lutsko, J.F.}
  \REVIEW{unpublished}{}{2009}{}.    
   
\bibitem{abe_99a}
  \Name{Abe, S.}
  \REVIEW{Phys. Lett. A }{263}{1999}{424}.     

\bibitem{abe_99b}
  \Name{Abe, S.}
  \REVIEW{Physica A }{269}{1999}{403}.
   
\bibitem{abe_09}
  \Name{Abe, S.}
  \REVIEW{Essential discreteness in generalized thermostatistics with
  non-logarithmic entropy}{}{2009}{Preprint}.



  
                  


\end{thebibliography}
\end{document}